\documentclass{elsart}

\usepackage{epsfig}
\usepackage{amssymb}

\begin{document}

\def\npdg{$\overrightarrow{n} + p \rightarrow d + \gamma$}

\begin{frontmatter}

\title{A Current Mode Detector Array for Gamma-Ray Asymmetry Measurements}

\author[LANL,IU]{M.T.~Gericke\corauthref{GSM}},
 \corauth[GSM]{Corresponding author. Tel.: + 1-505-665-7114.}
 \ead{mgericke@lanl.gov}
\author[LANL] {C.~Blessinger},
\author[LANL]{J.D.~Bowman},
\author[UManitoba]{R.C.~Gillis},
\author[IU]{J.~Hartfield},
\author[KEK]{T.~Ino},
\author[IU]{M.~Leuschner},
\author[KEK]{Y.~Masuda},
\author[LANL]{G.S.~Mitchell},
\author[KEK]{S.~Muto},
\author[IU]{H.~Nann},
\author[UManitoba]{S.A.~Page},
\author[LANL]{S.I.~Penttil{\"a}},
\author[TRIUMF,UManitoba]{W.D.~Ramsay},
\author[LANL]{P.-N.~Seo},
\author[IU]{W.M.~Snow},
\author[IU]{J.~Tasson},
\author[LANL]{W.S.~Wilburn}

\address[LANL]{Los Alamos National Laboratory, Los Alamos, New Mexico 87545, USA}
\address[UManitoba]{University of Manitoba, Winnipeg, Manitoba R3T 2N2, Canada}
\address[IU]{Indiana University, Bloomington, Indiana 47405, USA}
\address[KEK]{High Energy Accelerator Research Organization (KEK), Tukuba-shi, 305-0801, Japan}
\address[TRIUMF]{TRIUMF, 4004 Wesbrook Mall, Vancouver, British Columbia V6T 2A3, 
Canada}

\begin{abstract}
We have built a CsI(Tl) $\gamma$-ray detector array for the NPDGamma
experiment to search for a small parity-violating directional
asymmetry in the angular distribution of $2.2$~MeV $\gamma$-rays from the
capture of polarized cold neutrons by protons with a sensitivity of
several ppb. The weak pion-nucleon coupling constant can be determined
from this asymmetry. The small size of the asymmetry requires a high
cold neutron flux, control of systematic errors at the ppb level, and
the use of current mode $\gamma$-ray detection with vacuum photo diodes
and low-noise solid-state preamplifiers. The average detector
photoelectron yield was determined to be 1300 photoelectrons per
MeV. The RMS width seen in the measurement is therefore dominated by
the fluctuations in the number of $\gamma$ rays absorbed in the detector
(counting statistics) rather than the intrinsic detector noise. The
detectors were tested for noise performance, sensitivity to magnetic
fields, pedestal stability and cosmic background. False asymmetries
due to gain changes and electronic pickup in the detector system were
measured to be consistent with zero to an accuracy of $10^{-9}$ in a
few hours. We report on the design, operating criteria, and the
results of measurements performed to test the detector array.
\end{abstract}

\end{frontmatter}

\section{Introduction}
\label{scn:INSU}
Bright pulsed spallation neutron sources possess high instantaneous
neutron fluxes and also lend themselves to the measurement of neutron
energy by time of flight. New high precision fundamental neutron
physics experiments can be designed to take advantage of these
features~\cite{pp:FPPNB}. One such class of measurements consists of
searches for parity violation in polarized neutron capture on light
nuclei~\cite{pp:NIST1,pp:NIST2,pp:NIST3}.

NPDGamma, currently under commissioning at the Los Alamos Neutron
Science Center (LANSCE), is one such experiment. It is the first
experiment designed for the new pulsed cold neutron beam line, flight
path 12, at LANSCE. NPDGamma will determine the small weak
pion-nucleon coupling constant, $f_{\pi}$, in the N-N
interaction~\cite{pp:prop,pp:snow,pp:snow2}. This coupling constant is
directly proportional to the parity-violating up-down asymmetry,
$A_{\gamma}$, in the angular distribution of 2.2 MeV $\gamma$-rays with
respect to the neutron spin direction in the reaction \npdg,
\begin{equation}
  \frac{d\sigma}{d\Omega}\propto \frac{1}{4\pi}
  \left(1+A_{\gamma}\cos\theta\right).\label{eqn:CRSS}
\end{equation}
The asymmetry has a predicted size of $5 \times 10^{-8}$~\cite{pp:DDH}
and the goal of the NPDGamma collaboration is to measure it to 10\% of
this value.  The small size of the asymmetry imposes stringent
requirements on the performance of the beam line and apparatus. It is
necessary to achieve high counting statistics while at the same time
suppressing any systematic errors below the statistical limit.

The experiment makes use of an intense cold neutron beam at LANSCE
~\cite{pp:pil}.  The beam is pulsed at 20 Hz and transversely
polarized by transmission through a polarized $^{3}$He cell. A radio
frequency spin flipper is used to reverse the neutron spin direction
on a pulse-by-pulse basis. The neutrons are captured in a $20$ l
liquid para-hydrogen target. The 2.2 MeV $\gamma$-rays from the capture
reaction are detected by an array of 48 CsI(Tl) detectors. The entire
apparatus is located in a homogeneous $10$~G magnetic field to
maintain the neutron spin downstream of the polarizer and to suppress
Stern-Gerlach steering of the neutrons. Three $^{3}$He ion chambers
are used to monitor beam intensity, measure beam polarization and
transmission and monitor the ortho-para ratio in the liquid hydrogen
target.

To measure $A_{\gamma}$ to an accuracy of $5 \times 10^{-9}$, the
experiment must detect at least a few $\times 10^{17}$ $\gamma$-rays from
\npdg~capture with high efficiency. The average rate of $\gamma$-rays
deposited in the detectors for any reasonable run-time is therefore
high, and the instantaneous rates at a pulsed neutron source are, of
course, even higher than for a CW source. Because of these high rates
and for a number of other reasons discussed below, the detector array
uses accurate current mode $\gamma$ detection. Current mode detection is
performed by converting the scintillation light from CsI(Tl) detectors
to current signals using vacuum photo diodes (VPD), and the
photocurrents are converted to voltages and amplified by low-noise
solid-state electronics.

Another stringent constraint for the detector system is the detection
and elimination of any instrumental systematic effects inducing {\it
false} asymmetries associated with imperfections in the detector or
data acquisition (DAQ) system. These effects must be measured
periodically in the course of the experiment. It is therefore
essential to perform these measurements in a short time, compared to
the run time of the experiment. The time required for these
measurements is determined by the time required to average the
electronic noise.  For the current mode detection to be effective the
electrical noise in the detector system must be much smaller than the
beam-on shot noise.

A series of measurements have been performed both on individual
detectors and their components as well as on the detector array as a
whole, in conjunction with the DAQ.  The results show that the
detectors meet all requirements described above.  The remainder of
this paper describes the measurements in detail, including the setup,
the procedures used and the results found.  In particular, we report
on:
\begin{enumerate}
\item Section~\ref{scn:DETA}: Detector Design and Operational Criteria,
\item Section~\ref{scn:DETM}: Detector Photoelectron Yield and CsI to VPD Gain Matching,
\item Section~\ref{scn:ARRT}: Noise Performance, Background and False Asymmetry Studies,
\item Section~\ref{scn:CONC}: Summary.
\end {enumerate}

Several other characteristics of the detector array, such as long term
gain fluctuations and counting statistics performance, have to be
studied with a strong $\gamma$-ray source and are thus best done with
capture $\gamma$-rays in a cold neutron beam.  These tests have also been
conducted and will be discussed in a forthcoming paper.

\section{Detector Design and Operational Criteria} \label{scn:DETA}

\begin{figure}[h]
  \hspace{2cm}
  \includegraphics[bb=20 0 920 760,clip,scale=0.3]{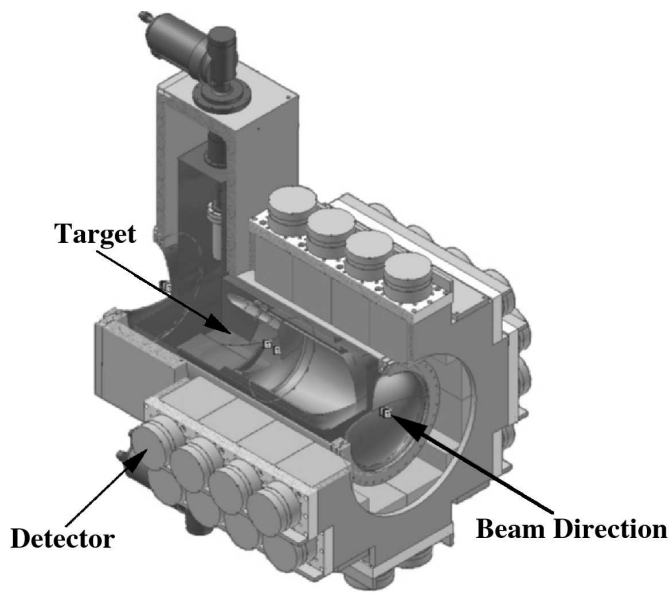}
  \caption{NPDGamma detector array and target assembly. The array surrounds a $20$~l
           liquid hydrogen target. There are 48 detectors grouped into four rings of 12 detectors 
           each and arranged in a cylindrical pattern around the hydrogen target.} \label{fig:DETA}
\end{figure}

The detector array consists of 48 CsI(Tl) cubes arranged in a 
cylindrical pattern in 4 rings of 12 detectors each around a
cylindrical $20$~l liquid hydrogen target (Fig.~\ref{fig:DETA}).
In addition to the conditions set on the detector array by the need to
preserve statistical accuracy and suppress systematic effects (see
Section~\ref{scn:SYSE}), the array was also designed to satisfy
criteria of sufficient spatial and angular resolution, high
efficiency, and large solid angle coverage. Here we discuss some of
the reasoning behind certain design choices and describe the specific
properties of our array.

To measure the asymmetry, a small ($5 \times 10^{-8}$)
parity-violating component must be detected in the presence of an
intense isotropic (parity-conserving) $\gamma$ signal. The parity-odd
component of the signal is proportional to $\cos{\theta}$, where
$\theta$ is the angle between the direction of neutron polarization
and the momentum vector of the emitted $\gamma$-ray. As long as the
change in $\cos{\theta}$ over a detector element is small, the finite
size of the detector elements will not reduce the statistical accuracy
of the experiment. From a calculation of the average $\cos{\theta}$
over the solid angle of a detector, the error in the measured
asymmetry due to spatial resolution, for $N$ detected $\gamma$-rays,
is $\sigma_{A_{\gamma}}=\sqrt{3}/\sqrt{N}$ for an infinitely fine
grained array and $\sigma_{A_{\gamma}}=2/\sqrt{N}$ for an array with
only two detectors, one covering each hemisphere~\cite{pp:prop}. Since
the error is a slowly-varying function of the degree of segmentation,
there is no pressing need for the detectors to be finely segmented,
and the lateral dimensions can be chosen with regard to other
criteria.

A segmented detector with elements large enough to fully contain the
$\gamma$ energy reduces the noise per detector from fluctuations in the
fraction of $\gamma$ energy shared among different detectors and
simplifies the identification of the $\gamma$ emission angle, which as
noted above need not be determined with high precision.  $\gamma$ cross
sections in high Z materials reach a minimum around $~2$ MeV energy
and the corresponding mean free path ($~5.5$~cm for $2.2$ MeV
$\gamma$-rays in CsI) sets the scale for the dimensions of the
detector elements. The size of the individual detectors is $152\times
152\times 152~{\rm mm^3}$. With these dimensions each crystal absorbs
about 84\% of the energy for a $2.2$ MeV $\gamma$-ray incident at the
center of the front face. MCNP\footnote{MCNP is a trademark of the 
Regents of the University of California, Los Alamos National Laboratory,
{\tt http://laws.lanl.gov/x5/MCNP/index.html}.}~\cite{pr:MCNP} and 
EGS4\footnote{The EGS code system and its various tools and utilities are
copyrighted jointly by Stanford University and National Research
Council of Canada. All rights reserved.
{\tt http://www.slac.stanford.edu/egs/}.}~\cite{pr:EGS4} calculations have
shown that about 3\% of the energy is backscattered from the front
face of the crystal, 11\% leaks out through the rear face, and 2\%
leaves through the 4 remaining sides. This reduces the cross talk
between detector elements to a level that is small enough to allow the
measurement of the asymmetry to the proposed accuracy. The main effect
of cross talk is a small loss in angular resolution.  A 20\% increase
in thickness in the direction of the incident $\gamma$-ray increases
the amount of energy absorption by only 4\%. 

The overall size of the array, on the other hand, is constrained by
the size of the source (target), which depends on the diameter of the
neutron beam $(10~{\rm cm})$ and the mean free path of a cold neutron in
the liquid hydrogen target ($14$~cm for a 2~meV neutron).  The liquid
hydrogen target is large enough to stop most of the neutron
beam. Monte Carlo calculations performed using the double differential
scattering cross sections for cold neutron scattering in liquid
parahydrogen~\cite{rep:Macfarlane} indicate that a $30$~cm diameter,
$30$~cm long target will capture about $60$\% of the incident
neutrons~\cite{pp:prop}. These calculations are based on the neutron
energy spectrum emitted by the coupled ${\rm LH_2}$ moderator viewed
by the NPDGamma beam line~\cite{pp:pil}. The neutrons that are not
captured in the liquid hydrogen are absorbed in a thin ($2$~mm)
plastic material loaded with $^{6}$Li to prevent activation of the CsI
by neutron capture. $\gamma$-Rays are transmitted through the low Z of the
plastic and the aluminum target vessel with high efficiency. Since
neutron absorption in $^{6}$Li is dominated by charged particle
emission as opposed to $\gamma$ emission, background $\gamma$-rays which can
dilute the signal from n-p capture are suppressed. For design purposes
we can therefore choose to concentrate on the signal from n-p capture
events rather than background events.

The detector array is arranged in cylindrical rings to surround most
of the hydrogen target and allow the neutron beam to enter and exit
without activating the CsI. It is important to detect the majority of
the photons emitted transverse to the neutron beam, since the neutrons
are transversely polarized and $\gamma$-rays emitted along the neutron
polarization direction contribute most to the parity-odd component of
the asymmetry. Photons emitted along the beam direction therefore
contribute little to the asymmetry, and the question becomes how many
rings need to be included. Monte Carlo calculations have shown that the
error in the asymmetry as a function of the number of rings along the
neutron beam axis reaches $87$\% of its asymptotic value for 4
rings~\cite{pp:prop}. Together with the individual detector dimensions
mentioned above this geometry covers a solid angle of nearly 3$\pi$.

With the detector size chosen to be large compared to the $\gamma$ mean
free path, the predicted peak $\gamma$ rate into a single detector in the
experiment is estimated to be $~100$~MHz, based on moderator
brightness measurements~\cite{pp:pil} and Monte Carlo calculations. At
this rate pulse counting is impractical for CsI given the decay time
of the scintillation light pulses (1
${\rm{\mu}s}$~\cite{pp:Grassman,pp:Schotanus}).  For an array composed
of CsI detectors, the high photon rates must be handled using current
mode $\gamma$ detection.

The crystals were manufactured and encased in the housing by Bicron
\footnote{Bicron, Saint-Gobian Industrial Ceramics, Inc.
{\tt www.bicron.com}}. Each detector module consists of two rectangular
pieces of optically coupled Thallium doped Cesium Iodide crystals. The
slightly hygroscopic CsI(Tl) crystals are wrapped in PTFE Teflon, a
diffuse reflector, and hermetically sealed in a $1.0~{\rm mm}$ thick
Aluminum housing. The {\it Optics} program~\cite{pp:FRLEZ} was used to
study which reflector and crystal surface treatment to use, in order
to obtain the maximum light output and best overall uniformity for the
given detector geometry. We found that diffuse reflection produced the
best results. The crystals are coupled to a $76~{\rm mm}$ diameter K+
glass window at the top of the housing assembly to facilitate the
detection of the scintillation light by a vacuum photodiode (VPD)
during standard operation. The detectors are individually mounted on
the array to minimize potential stress and plastic deformation of the
crystals.

\begin{figure}[h]
  \hspace{3cm}
  \includegraphics[scale=0.5]{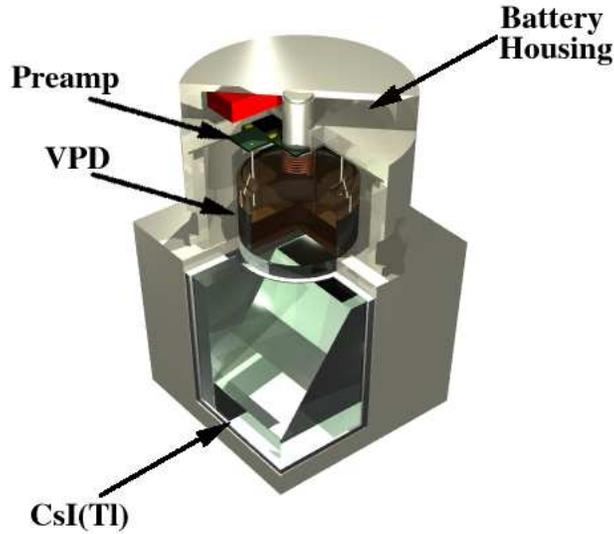}
  \caption{Illustration of an individual detector. Each detector consists of two coupled CsI(Tl)
           scintillators, a VPD and a preamplifier stage.} \label{fig:IDET}
\end{figure}

CsI(Tl) was chosen because of its high density ($4.53~{\rm g/cm^3}$),
large {\it Z}, high light yield and its relatively low cost. For the
alkali iodides, thallium activation is required to achieve a high
light output. For CsI(Tl) the emitted scintillation light has a
wavelength centered around $540~{\rm nm}$ which is well matched with
the absorption characteristics of the type of VPD used in this
experiment.  The light yield of CsI(Tl) is 54000 photons per MeV at
maximum emission and with a scintillation efficiency of
12\%~\cite{bb:Knoll}.  The current collected from the VPD anode is
amplified by a low noise solid-state amplifier.

A cylindrical aluminum housing for the VPD and preamplifier is mounted
on top of the CsI crystal assembly. The housing is designed to be
light-tight, to minimize noise contributions from capacitive coupling
between electronic components on the preamplifier board and to shield
the assembly from outside fields such as those produced by the radio
frequency spin flipper. To avoid ground loops the detector housing is
grounded via the signal cable shield only and the detectors are
individually mounted to a stand which allows the electrical isolation
between detectors.  Each detector comes equipped with two light
emitting diodes (LED), one in each crystal half. The LEDs are used
during beam off detector diagnostic tests (see Section~\ref{scn:FAM}).

Radiation damage will decrease the self-transparency of the crystals,
resulting in a decrease in detected light. CsI(Tl) has been found to
be rather radiation hard up to doses of more than
$500$~Gy~\cite{pp:REN,pp:ZHU,pp:WEI,pp:CHD,pp:KOB}, with the precise
threshold for significant radiation damage in the crystal dependent on
crystal impurities as well as the radiation damage rate. This
radiation dose is approximately the dose that the detectors will
receive over the course of the entire experiment (a few thousand hours
of running), and the corresponding damage rate is small compared to
those that have caused significant radiation damage in CsI(Tl)
detectors in the past.

\subsection{Vacuum Photodiodes} \label{scn:VPDI}

To convert the scintillation light to a current the detectors employ
$76$~mm S-20 Hamamatsu
\footnote{ VPD type R2046PT, Hamamatsu Corporation, 360 Foothill Road,
P.O. Box 6910, Bridgewater, N.J. 08807-0910, USA, www.hamamatsu.com}
vacuum photodiodes rather than photomultiplier tubes (PMT).  The
decision to use VPDs was based on the fact that photomultipliers are
very sensitive to magnetic fields. A $1$~G field leaking into the PMT
can change its gain by 100\%.  The experiment uses magnetic fields to
control the neutron spin direction and any field leaking into the
detectors may produce large gain changes. On the other hand, the
sensitivity of a vacuum photodiode to magnetic fields is only about
$1\times 10^{-4}{\rm{/G}}$ in a $10$~G DC field or $1\times
10^{-5}{\rm{/G^2}}$ for $10$~G AC field (see Section~\ref{scn:LVT})
~\cite{pp:greg}.

This particular type of vacuum-photodiode was chosen for the low
photocathode sheet resistivity. The low sheet resistivity of the S-20
photocathode reduces the degree of gain non-linearities across its
surface.With an S-20 cathode, the VPD has a quantum efficiency of
$\approx 10\%$ at the CsI maximum emission wavelength.  A bias of
$90$~Volts is applied across the VPD via two $45$~V batteries located
on top of the VPD and preamplifier housing (Fig.~\ref{fig:IDET}). This
removes the necessity for an external supply to be connected directly
to the VPDs which could cause ground loops and introduce additional
noise in the VPDs. The batteries
\footnote{Eveready Industries India, LTD. ,
www.evereadyindustries.com} have a capacity of $140$~mAh, the average
beam-on current drawn from the VPDs is about $30$~nA and the VPDs can
therefore nominally be run for $10^7$~hours continuously.  So the time
before exchange should be limited by the battery's five year shelf
life.

Photodiodes are known to be extremely linear devices. Bench tests
with the photodiodes used in the experiment have shown that their
gain is uniform to better than $2\times 10^{-6}$~per nA of photocathode
current up to 500 nA. The typical peak photocathode current per detector
for the NPDGamma experiment is $\sim 50$~nA, while pulse-to-pulse fluctuations
seen by the detectors are typically less than 1\%.

\subsection{Low Noise Preamplifier} \label{scn:AMPI}

\begin{figure}[h]
  \hspace{1.0cm}
  \includegraphics[scale=0.5]{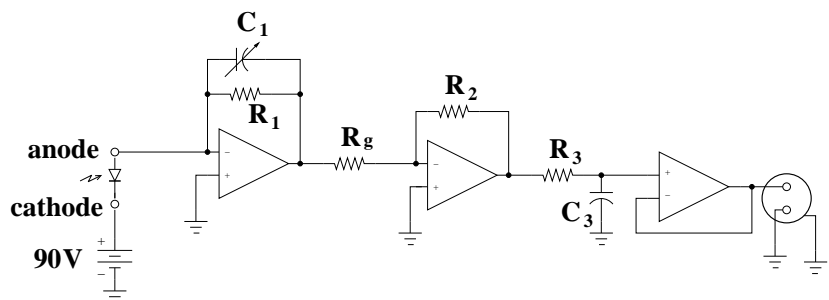}
  \caption{The detector preamplifier consists of a decoupled power supply,
           two $45$~V batteries to provide the bias across the VPD and
           three operational amplifier stages.} \label{fig:PREAMP}
\end{figure}

The VPD current in each detector is converted to a voltage signal by a
three stage low-noise current-to-voltage preamplifier~\cite{pp:swb}.
The first stage uses an op amp-based current-to-voltage amplifier with
a gain of $5\times 10^{7}$ (Fig.~\ref{fig:PREAMP}). The second
amplifier stage serves as an inverter with a nominal gain of
$-2.15$. The resistors in the second stage are used to adjust relative
detector gains (see Section~\ref{scn:DAQ} and~\ref{scn:CRGT}).  The
third stage serves as a low impedance line driver.

The detector preamplifier has been designed to operate at noise levels
close to the theoretical limits set by Johnson noise, so that the time
required to measure the asymmetry to the level of $5\times10^{-9}$ is
dominated by the collection of counting statistics rather than by the
need to average electronic noise~\cite{pp:pmsn}. Various filter stages
and ground isolation between the preamp power and signal circuit have
been implemented to satisfy these requirements.

\subsection{Data Acquisition and Storage}\label{scn:DAQ}

Because of the small size of the parity violating signal and the
presence of the large isotropic signal, the $\gamma$ intensities within
each of the 8 near detectors and each of the 4 corner detectors in a
given ring are nearly equal. It is therefore possible to equalize the
signals from near and corner detectors by gain adjustments (discussed
below) and sample only (1) the average signal in a ring, (2) the
differences in each detector from the average signal in a ring. This
strategy allows one to exploit the increased dynamic range to increase
the gain of the system and minimize the effects of noise in the
sampling electronics.  Here we describe the details of how this idea
for sampling the array signals was implemented.

\begin{figure}[h]
  \hspace{3cm}
  \includegraphics[scale=0.3]{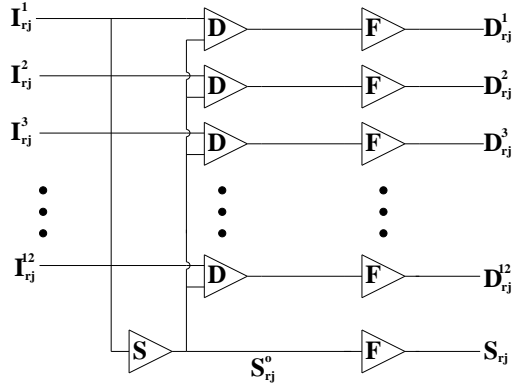}
  \caption{DAQ sum and difference amplifier schematic. The 12 individual detector
           signals for the ring are denoted by $I^i_{rj}$. The 12 corresponding
           difference signals are denoted by $D^i_{rj}$. The ring average signal
           is denoted by $S_{rj}$. Here, $r,j$ denotes the {\it rth} ring and the 
	   {\it jth} sample.} \label{fig:SDAS}
\end{figure}

\begin{figure}[h]
  \hspace{2.0cm}
  \includegraphics[bb=100 0 575 766,clip,angle=270,scale=0.35]{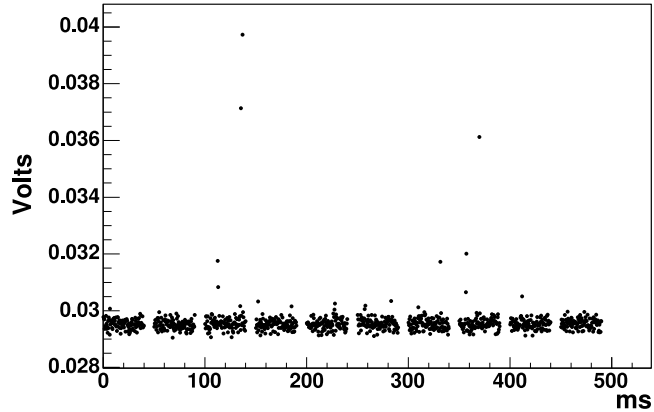}
  \caption{A typical detector pedestal signal, showing 10 macro pulses
           with 100 data points each. Each time bin is $0.4$~ms wide
           containing one data point. The $40$~ms long sampling period
           is followed by a $10$~ms break in each pulse. Also seen are
           about 14 outliers, where the larger signals are due to
           incident cosmic rays depositing up to$\sim 100$~MeV in the
           crystal.} \label{fig:DOUT}
\end{figure}

The preamplifier output for each detector is sampled by the NPDGamma
data acquisition. The DAQ incorporates four sum and difference
amplifier boards with 12 channels each. Each sum and difference board
forms an average voltage over the 12 detectors in a given ring and
each individual detector signal has its corresponding ring average
subtracted. The process is shown schematically in
Fig.~\ref{fig:SDAS}. Here each difference amplifier contributes a gain
factor of 10 and each Bessel filter (denoted by F in the schematic)
contributes an additional factor of 3 to the gain. The 48 resulting
difference signals and four average signals are sampled by 16-bit
ADCs. The sum and difference signals are sampled at $62.5$~kHz and
$50$~kHz respectively.  A {\it macro pulse} of data is collected by
sampling for a duration of $40$~ms, followed by a $10$~ms break,
before the next frame of neutrons arrives. This results in 2000
difference and 2500 sum samples for each macro pulse.  In the data
stream (before the raw data are written to file) every group of 20
difference samples and 25 sum samples is summed to produce a final
value for each of 100, $0.4$~ms wide time bins. The sampled data are
transfered via fiber optic connection to a 3.5 Tbyte RAID array
storage device. Figure~\ref{fig:DOUT} shows the 10 macro pulses of
electronic pedestal (beam off) output for a typical detector, obtained
using the described sampling scheme.

According to Fig.~\ref{fig:SDAS} and the sampling scheme just
described, the data actually stored for each time bin are a sum of 20
difference samples for each detector ${\mathcal D}^i_r =
\sum^{20}_{j=1}D^i_{rj}$ and a sum of 25 average samples for each ring
${\mathcal S}_r = \sum^{25}_{j=1}S_{rj}$. A ring average sample is
given by $S_{rj} = 3/12~\sum^{12}_{i=1}I^i_{rj}$ and a difference
sample for a given detector in the ring is given by $D^i_{rj} =
30(I^i_{rj} - S_{rj}/3)$.  In the analysis, the time bin average of
the difference and sum signals are recombined to produce the average
detector signal for the time bin at the ADC input $\overline{I^i_r} =
1/30~(\overline{D^i_r} + 10~\overline{S_r})$, in ADC counts. Here,
$\overline{D^i_r} = {\mathcal D}^i_r/20$ and $\overline{S_r} =
{\mathcal S}_r/25$.

The sum and difference scheme increases the effective dynamic range of
the ADCs, which are limited to $\pm 10$~V, therefore allowing for a
larger gain to be applied and staying above the bit-noise of the
ADCs. The Bessel filters provide highly correlated ADC samples,
filtering out high frequency components in the signal, and the high
sampling rate averages out the bit noise in the ADCs. The chosen time
bin width removes the correlation between the data points actually
used in the calculation of asymmetries (see Section~\ref{scn:DNP}).

\subsection{Mode of Operation and Systematic Effects}\label{scn:SYSE}
Achieving the desired accuracy in measuring the parity violating $\gamma$
asymmetry depends on good counting statistics with comparatively small
errors from other sources, such as electronic noise and systematic
effects.  During beam-off measurements, the time required to determine
any systematic effect is governed by the noise in the preamplifier and
the rest of the DAQ.  Since these effects need to be studied
periodically during the experiment, it is essential to perform the
beam-off measurements quickly compared to the time required to collect
counting statistics. To satisfy these requirements, the detector array
must have a high photoelectron yield, a low sensitivity to external
radioactivity and electromagnetic effects and very good noise
performance.

The photoelectron yield enters into the calculation of the average
photo current seen at the detector preamplifier output as well as the
shot-noise seen at the VPD cathode. The time needed to measure an
asymmetry to a given accuracy is proportional to the inverse of the
average photo-current.  In a current mode measurement, if the detector
has a high photoelectron yield, counting statistics are manifest in
the form of shot noise due to the fluctuations in the number of $\gamma$-rays
entering the detector. The corresponding expected RMS width is given
by~\cite{bb:DVR,bb:HWHL}
\begin{equation}
 \sigma_{I_{\mathrm{shot}}} = \sqrt{2qI}~\sqrt{f_B}, 
\end{equation}
where $q$ is the amount of charge created at the photo cathode per
detected $\gamma$-ray, $I$ is the average photo-current per detector and
$f_B$ is the frequency bandwidth, set by the filtering in the data
acquisition system (DAQ).

Since the experiment intends to determine an asymmetry with a
precision of $5\times10^{-9}$, any systematic effect resulting in a
false asymmetry has to be measured to at least this level of accuracy
in a short period of time. The measurement of such a small quantity
requires the careful evaluation and analysis of any possible
systematic effects. For the detector array the two most serious
potential instrumental systematic effects may be caused by a radio
frequency spin flipper (RFSF)~\cite{pp:greg}, which is used to reverse
the spin of the neutrons. The RFSF is a $30$~cm diameter and $30$~cm
long solenoid enclosed in an aluminum housing. It operates according
to the principles of NMR, using a $30$~kHz magnetic field with an
amplitude of a few G and will be mounted partially inside the detector
array. The neutron spin direction is reversed when the RFSF is on and
is unaffected when it is off. To turn the RFSF off, the current drawn
by the coils is switched to a dummy load consisting of a resistor
circuit designed to have the same impedance as the coils. This keeps
the load on the main power circuit constant and minimizes pickup of
the spin flipper on-off switching in other circuits.

During normal (beam-on) operation, when $\gamma$-rays from neutron capture
create a large signal in the array, any magnetic fields leaking into
the VPDs can produce a systematic effect through a multiplication of
the overall detector gain. We call such an effect a multiplicative
systematic error. In addition, any electronic pickup could add a false
signal on top of the real signal. We call such an effect an additive
systematic error. If these signals are correlated with the spin state
of the neutrons, through the spin flipper, this could lead to false
asymmetries.

The efficiency of the $\gamma$-ray detectors will change slowly due to a
number of effects. The primary technique for reducing false
asymmetries generated by these slow changes is fast neutron spin
reversal. This allows asymmetry measurements to be made for opposing
detectors for each spin state and very close together in time, before
significant drift occurs. Note that the asymmetry is measured
continuously since the signals from opposite detectors are measured
simultaneously for each spin state.  By carefully choosing the
sequence of spin reversal, the effects of drifts up to second order
are further reduced (See Section~\ref{scn:ASYDEF}).

\section{Detector Photoelectron Yield and CsI to VPD Gain Matching}\label{scn:DETM}
There are important reasons for making the overall efficiency of the
elements of the detector array as uniform as possible. Equalizing the
overall efficiencies through relative gain matching between detectors
prevents saturation of the difference signal channels in the ADC (see
Section~\ref{scn:DAQ}) and allows an expansion of the dynamic range as
discussed earlier.  Furthermore, uniform detector efficiencies make
the observed parity-odd up-down $\gamma$ asymmetry signal less sensitive
to potential neutron spin-dependent crosstalk from parity-conserving
left-right asymmetries which are known to be present at small levels
in the interaction of the neutrons with hydrogen and in the $\gamma$
angular distribution~\cite{pp:CGP}. Unfortunately it was not practical
to obtain all the individual components of the detector with
sufficiently uniform properties to ensure this by design. For these
reasons great care has been taken to characterize the relevant
properties of all of the individual components for each detector so
that they can be individually matched to minimize variations in the
overall gain of each detector/VPD/preamp combination.  After hardware
matching is optimized, final adjustments can be made in amplifier
gains and also in software. This section describes these measurements.

To establish the properties and performance of the individual detector
components, a variety of measurements were performed prior to their
assembly.  The photoelectron yield of the CsI scintillators and the
efficiency of the VPDs were measured independently and the results
were used to match them and obtain a reasonably uniform relative gain
between all CsI-VPD detector modules. After assembly of the detectors,
a current mode measurement was performed to establish the combined
detector gain and to refine the gain matching using the resistors in
the second preamplifier stage. 

\subsection{CsI Relative Photoelectron Yield}\label{scn:RGE}

The primary photo-peaks of two radioactive sources, $^{241}{\rm
Am}~(0.4~{\rm MBq},~E_{\gamma} = 0.06~{\rm MeV})$ and $^{137}{\rm
Cs}~(0.3~{\rm MBq},~E_{\gamma} = 0.67~{\rm MeV})$ and standard pulse
counting methods were applied to determine the number of
photoelectrons per MeV
from the RMS width of their respective peaks. This procedure relies on
the assumption that the photo-peak widths (${\sigma}_p = {\rm
FWHM}/2.35$) are due primarily to the fluctuations in the number of
photoelectrons made at the photocathode and subsequent dynodes of the
PMT (shot noise), as well as intrinsic properties of the
crystal. Contributions to the peak width due to electronic noise were
combined with those due to crystal intrinsic properties into a single
width $\sigma_{int}$.

\begin{figure}[h]
  \hspace{4cm}
  \includegraphics[scale=0.55]{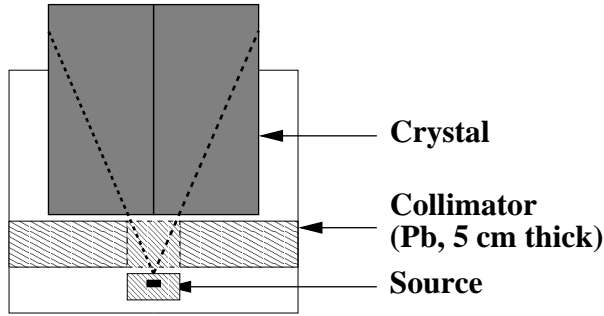}
  \caption{Setup for detector photoelectron yield measurement. The $\gamma$-ray sources 
    were centered such that both scintillator crystals are illuminated equally. 
    The sources were collimated to avoid smaller energy deposits in partially contained
    events.} \label{fig:PHS}
\end{figure}

The $\gamma$ sources were centered on one side about $6.8~{\rm cm}$ from
the detector housing to illuminate both halves of the crystal equally
(Fig~\ref{fig:PHS}). The $\gamma$-rays were collimated down to $\sim
0.2~{\rm sr}$, using a $5~{\rm cm}$ thick lead shield. The source was
mounted on a reproducible mount, to ensure that the relative
source-detector position was always the same. A $127~{\rm mm}$
Hamamatsu R1513 PMT was optically coupled to the detector window using
BCS 260 optical coupling grease. To study the effects of optical
coupling quality on the overall detector efficiency two randomly
chosen detectors were later used in conjunction with two other
coupling methods, and the results were compared to those obtained here
(see Section~\ref{scn:CVM}).

To extract the photoelectron yield, a plot of the two relative peak
variances $\Sigma^2 = \sigma^2_p / \overline{p}^2$ versus inverse
peak-energy was made. A linear fit was made between the two points,
using $$\Sigma^2 = a\frac{1}{E} + \sigma_{int}^2.$$ Here
$\overline{p}$ is the zero-offset corrected peak mean. The slope of
the line $$a=\frac{1}{N}(1+\frac{1}{\delta-1})$$ was extracted from
the fit and determines the number of photoelectrons per MeV
$$N = a^{-1}\frac{\delta}{\delta-1} \simeq \frac{1.37}{a}.$$ Here,
$\delta$ is the Poisson distributed gain in the number of electrons
produced at the PMT cathode which, for a $10$ stage Hamamatsu R1513
PMT operated at $1.6$~kV, is $\sim 3.7$. The overall gain for the PMT
used is $5 \times 10^{5}$ at $1.6$~kV.  The factor $\frac{1}{\delta-1}
\simeq 0.37$ emerges due to the fluctuations in the number of
electrons made at each dynode stage of the PMT, which contribute to
the overall RMS width in the photopeaks. Corresponding to the
activities of the two sources, the error in the yield due to counting
statistics for the time counted are $\sim 0.1$\% for $^{241}{\rm Am}$
and $\sim 0.07$\% for $^{137}{\rm Cs}$. The overall error on the
results is dominated by the quality of the least-squares fit.

\begin{figure}[h]
  \hspace{2cm}
  \includegraphics[scale=0.45]{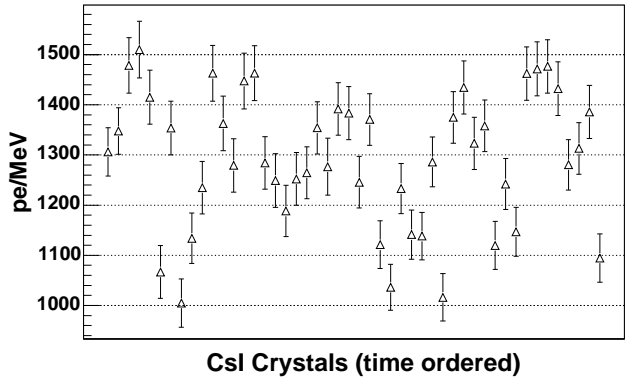}
  \caption{CsI detector efficiency. The average measured efficiency is 
           $1300 \pm 260$ photoelectrons (pe) per MeV. The efficiencies are
          shown in the order they were measured.} \label{fig:CSIEFF}
\end{figure}

Possible gain nonuniformities from the two crystal halves were
searched for by separately exposing each half to the $^{137}{\rm Cs}$
source.  Measurements were taken with the source on the left side,
right side and the center of the scintillator, where the center is
defined as seen in Fig.~\ref{fig:PHS}.  In the worst case, the gain
varies by about 7\% from one crystal half to the other.  The
photoelectron yield measurements produced an average of 1300
photoelectrons per MeV with an overall variation of $\pm~20$\% between
detectors. Figure~\ref{fig:CSIEFF} shows the results for 48 detectors
in the order they were taken. The error in the values is about $5$\%,
mostly due to the fitting procedure.

\subsection{VPD Relative Gain and Efficiency} \label{scn:LVT}
The response of the VPDs to the CsI scintillation is different from
the response seen using a tungsten lamp, which was used by Hamamatsu
Photonics to calibrate the VPDs. Therefore, it was decided that the
VPD relative efficiencies had to be studied using the CsI
scintillation light. The VPD efficiencies were measured using capture
$\gamma$-rays from a neutron beam at KEK and relative efficiency measurements
were performed at Los Alamos, using the LED's in the detectors (see
Section~\ref{scn:LVT}). At the pulsed epithermal neutron beam line at
KEK, Cd and In targets were used to convert neutrons to $\gamma$-rays
through radiative neutron capture at the Cd cutoff and at the In
$1.46~\rm{eV}$ and $9.1~\rm{eV}$ resonances. One of the CsI crystals
and its preamplifier were installed, together with each tested VPD,
next to the target, using proper neutron and $\gamma$ shielding. In this
way, relative efficiencies of 57 VPDs were determined with an error of
6\%. Comparisons of normalized VPD efficiencies at different neutron
energies show a very high correlation, whereas the correlation between
efficiencies measured with neutrons and those measured with a tungsten
lamp is very poor~\cite{pc:Ino}.

\begin{figure}[h]
  \hspace{3cm}
  \includegraphics[scale=0.45]{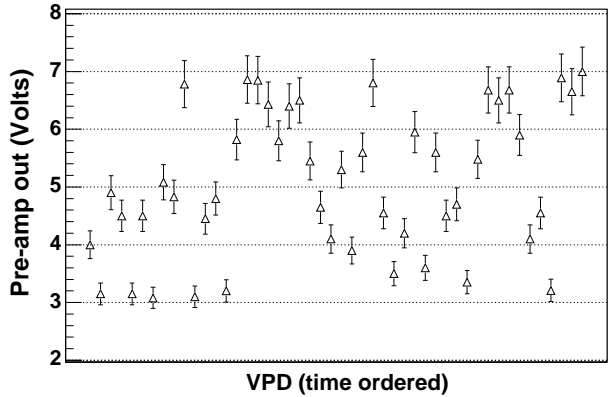}
  \caption{VPD relative efficiency results using LEDs. The efficiencies are
           shown in the order they were measured.} \label{fig:VPDEFF}
 \end{figure}

In a separate measurement the LEDs in each detector were used to
establish the relative VPD efficiency again and verify the quality of
the CsI-VPD gain matching.  To measure the relative VPD efficiency, a
single CsI crystal was coupled, in turn, with each VPD, using vacuum
grease as coupling compound (see Section~\ref{scn:CVM}). A $2$~V,
$100$~Hz square wave was applied to both LEDs in the detector.  The
current drawn by the LEDs was constant at $~18.5$~mA throughout the
measurement for 48 VPDs.  The preamplifier output was monitored with a
scope and with a precision voltmeter.  The relative gains differ by up
to a factor of 2.5. The results are shown in Fig.~\ref{fig:VPDEFF}.  A
conservative estimate on the error in the efficiency is $~7$\%, based
on the fluctuations seen in the output of the precision voltmeter.

\begin{figure}[h]
  \hspace{3cm}
  \includegraphics[scale=0.45]{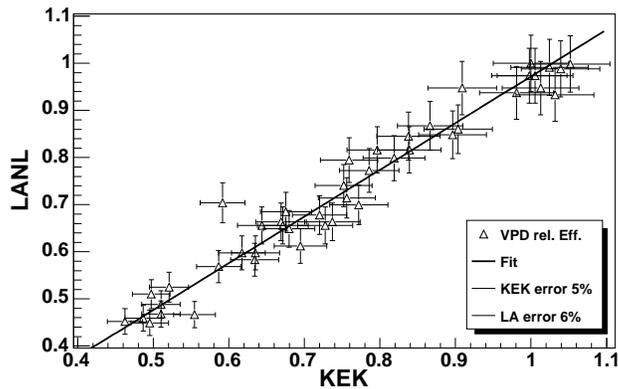}
  \caption{A comparison of VPD relative efficiencies, measured with neutrons at KEK and
           with LEDs (see text).} \label{fig:VPDLAKEK}
 \end{figure}

These measurements are compared with those done at KEK.
Figure~\ref{fig:VPDLAKEK} shows that the results of the two
independent measurements agree to within errors. The differences are
most likely due to the change in optical coupling when switching the
VPDs.

\subsection{CsI to VPD Matching} \label{scn:CVM}

As already mentioned, the CsI and VPD efficiencies are matched to
reduce the overall gain variations in the detector array. However,
fluctuations in the quality of the VPD-CsI optical coupling also
affect the overall gain. A pair of detectors was tested with three
optical coupling compounds:
\begin{enumerate}
\item BC 630 Bicron optical coupling grease,
\item Dow Corning Sylgard 184 Silicone Elastomer (cookies) and
\item high vacuum grease (translucent).
\end {enumerate} 

The Sylgard elastomer was used to make ``cookies'', circular $76$~mm
diameter disks, about 3 mm thick. The initially liquid compound was
pumped to remove air and cast into a mold standing on its side to make
the surfaces as flat as possible.  Unfortunately the compound was too
hard and small nonuniformities on the VPD window or on the casting
surfaces prevented the elastomer from producing a significant increase
in coupling quality. The best results for the Sylgard elastomer are
achieved by pouring it and allowing it to set in place, between
scintillator and VPD. However, this option was not considered because
the detector array configuration makes it difficult to exchange entire
CsI-VPD assemblies {\it in situ}. The best efficiency results were
obtained using optical coupling grease, which increased the gain by a
factor of two over no coupling compound. However, BC 630 was
disqualified due to its low viscosity and the vertical orientation of
some of the VPD-CsI boundaries in the array. Since high vacuum grease
is much more viscous and gave $90$\% of the gain of BC 630, it was
chosen for the coupling.

\begin{figure}[h]
  \hspace{3cm}
  \includegraphics[scale=0.45]{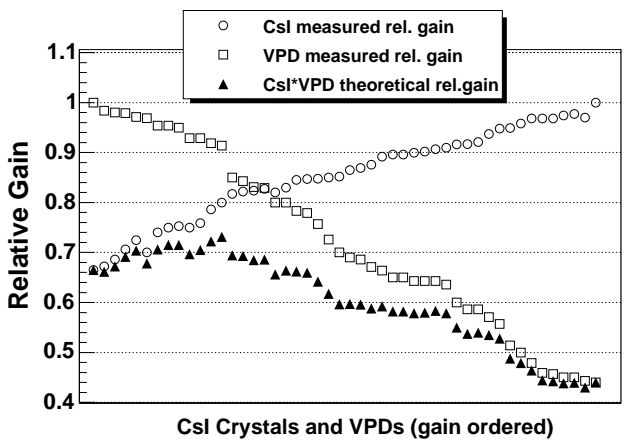}
  \caption{CsI and VPD gain matching. The CsI relative gain is shown in 
           increasing order and the VPD relative gain is shown in decreasing
           order. Their product shows the theoretical overall efficiency 
           spread after matching the VPDs and scintillators in this way.} \label{fig:CSIVPDGM}
\end{figure}

Figure~\ref{fig:CSIVPDGM} shows the gain ordered efficiency
measurement results for the VPDs, the CsI crystals and their
product. The curve showing the product of the efficiencies provides an
upper limit on how well the gain of the detectors can be matched in
hardware without further adjustments. The actual relative gain shifts
are expected to differ from this prediction, due to variations in
optical coupling quality among the detectors and variations in
scintillation light response of the crystals over time. The CsI
crystals were matched with the VPDs according to the efficiency
results presented in Fig.~\ref{fig:CSIVPDGM}. This selection served as
a starting point from which to conduct additional efficiency
measurements and further improve the gain via the adjustment of the
feedback resistors in the detector preamplifier.

At this stage, all feedback resistance values in the preamplifiers
(Fig.~\ref{fig:PREAMP}) were nominally identical for all
detectors. The overall variation observed in the efficiency of the
completely assembled detectors determine the change in feedback
resistance necessary to adjust the gain in any given detector.

\subsection{Combined Relative Detector Gain} \label{scn:CRGT}

To compare the predicted values of the relative gain between
detectors, the assembled detectors were tested with a $^{137}$Cs
source, intense enough to produce a current mode output. The
corresponding preamplifier output was fed through a low-pass filter
and monitored with a precision voltmeter. The source was located flush
with the outside surface of the detectors and centered on one side. A
source with the given activity (see Section~\ref{scn:RGE}) and in the
given configuration (fractional solid angle $\simeq 0.4)$ is expected
to deposit approximately $6.54\times10^{4}~{\rm MeV/s}$ into the
detector.  With the measured average CsI photoelectron yield of $\sim
1300~{\rm pe/MeV}$ one expects to measure an output on the order of a
few millivolts.

\begin{figure}[h]
  \hspace{3cm}
  \includegraphics[scale=0.45]{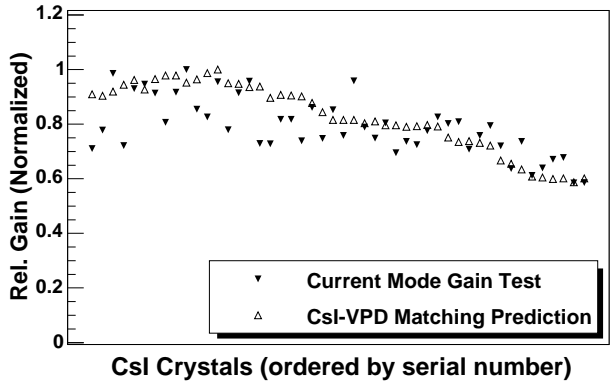}
  \caption{Detector relative gain measured in current mode, as compared to the
           theoretical prediction. The gains were measured on the bench under
           identical conditions for each detector.} \label{fig:CMABSGT}
 \end{figure}

The low-pass filter was adjusted to have a time constant of 15 seconds
to stabilize the voltage. Three measurements were taken for each
detector, one without source, one with source in place, and again
without the source. The measurements were performed quickly to avoid
fluctuations with long time constants ($\sim$ minutes). The mean
voltage out of the preamplifier was 2.4 mV. Figure~\ref{fig:CMABSGT}
shows the normalized difference between the signal with source and the
average of the two signals taken without the source, versus detector
number. The overall spread in gains is still $\sim 40$\%, but the
relative gains shifted a bit, as expected.

\begin{figure}[h]
  \hspace{3cm}
  \includegraphics[scale=0.45]{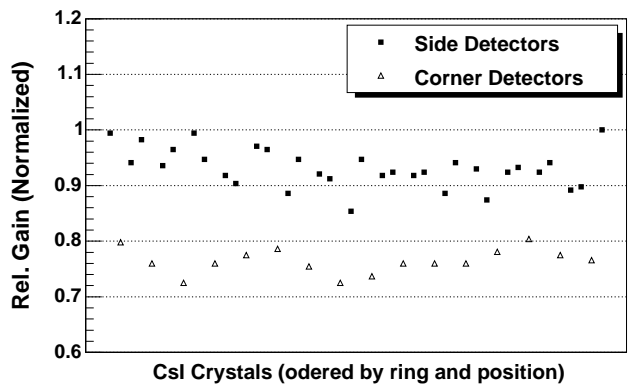}
  \caption{After assembly of the detector array the relative gain was again
           measured using a rotating $\gamma$ source located at the center of
           each detector ring. The corner detectors are shadowed by the detectors
           above an below them and therefore cover a smaller solid angle.} \label{fig:LOCKIN}
 \end{figure}

Resistors ranging from $1.50~{\rm k\Omega}$ to $2.80~{\rm k\Omega}$
were then installed in the second amplifier stage to reduce the gain
variations. After adjusting the preamplifier resistors and assembling
the detectors into the array stand (Fig.~\ref{fig:DETA} for the final
configuration), an additional set of data was taken to verify the
final relative detector gains (before using the gain modules). This
measurement was done, using a rotating $2~{\rm MBq}$ $^{137}$Cs source
and a lock in-amplifier. The source was located at the center of the
ring corresponding to the detector that was tested. The reference
phase for the lock in-amplifier was generated by an infrared
emitter-receiver mounted onto the shaft of the rotating source. The
measured normalized relative gains per detector are shown in
Fig.~\ref{fig:LOCKIN}. The lock in-amplifier and the rotating source
were used to filter out noise and other fluctuations to obtain a
cleaner measurement of the detector gains.

The corner detectors cover a solid angle that is $\sim 20$\% smaller
than it is for the side detectors (Fig~\ref{fig:DETA} and
Fig~\ref{fig:RING}). The gain in the corner detectors was then 
matched with the rest of the array by further adjusting their feedback
resistors. This level of hardware gain matching is sufficient to
prevent the difference channels from saturating.

The final precision of gain matching in the array will be limited by
the quality of optical coupling as well as any time dependent gain
fluctuations in the detectors. The next stage of gain matching will be
implemented using custom built adjustable gain VME modules in the DAQ
stream and will be performed during the commissioning run, in
conjunction with a neutron beam and signals from neutron capture on a
target. This work will be described elsewhere.

\section{Noise Performance, Background and False Asymmetry Studies}~\label{scn:ARRT}

In this section we describe studies of long term fluctuations in
detector pedestals and electronic noise due to radioactive and
electromagnetic background. We show that cosmic ray background in the
detector array is understood and has no effect on the measured
asymmetries. We discuss measurements performed to verify that false
asymmetries due to electronic pickup and magnetic field induced gain
changes in the VPDs are negligible.

These measurements required the acquisition of data over long periods
of time and under conditions that are as close as possible to those
encountered when the experiment is running. Accordingly, the entire
array and spin flipper were assembled in their final configuration and
data have been taken with the same DAQ setup to be used in the final
experiment.

\subsection{Detector Noise}\label{scn:DNP}

The accuracy of the measurements described in this paper has to be
viewed in relation to the noise levels in the detector pedestals.

In a current mode experiment, the accuracy of the measurement is
governed by the rate and quality of sampling of a signal that may be
viewed as a continuous string of values of a random variable. The
randomness and the spread (RMS width) of the samples determine how
many samples one must take to achieve a certain level of accuracy in
the measurement, while the sampling rate determines how long that will
take and whether or not one has in fact measured a representative
subset of the signal. As the sampling rate is increased, a larger
fraction of the width in the signal is due to the correlation between
samples, and the observed error will be larger than expected from
$1/\sqrt{N}$ counting statistics.  Thus, for a given number of samples
taken, oversampling leads to loss in statistical
information. Undersampling, on the other hand, will lead to
information loss in the noise since high frequency fluctuations are
aliased into lower frequencies.

The detector preamplifier (see section~\ref{scn:AMPI}) and the rest of
the DAQ were designed under the requirement that there be no
substantial additional noise contribution beyond the expected Johnson
noise from the resistors in the first preamplifier stage and the
intrinsic noise of the operational amplifiers used~\cite{pp:swb}.  To
verify that this constraint has been met, it is important to consider
the noise behavior one would expect based on the circuit design.

\subsubsection{Expected Noise Performance}~\label{scn:ENP}

To establish the actual RMS width one expects to see in the voltage
signal at the output of the preamplifier, one has to understand the
origin of all noise contributions, their propagation through the DAQ
and their final processing in the data stream.  This includes
determining a suitable sampling rate and the correct bandwidth for the
predicted noise levels, set by the filtering in the DAQ.

For the purposes of noise analysis, the most important component of
the preamplifier circuit (Fig~\ref{fig:PREAMP}) is the first stage
connected to the VPD anode. This stage incorporates a $50~{\rm
M\Omega}$ feedback resistor (${\rm R_1}$) which is expected to
completely dominate the noise. Subsequent resistors in the
preamplifier and DAQ are smaller by three orders of magnitude and
their noise contribution is thus negligible. However, each additional
amplifier and filter stage will have a multiplicative effect on the
noise generated in the first stage.

The thermal noise spectral density in the output of the first
amplifier stage is predicted to be
\begin{equation}
 \mathcal{S}_{V_J}(f) = 4{\rm k_B}TR_1~~{\rm \left[ \frac{V^2}{Hz} \right]}. \label{eqn:JNN}
\end{equation}
Here $k_B$ is Boltzmann's constant and $T$ is temperature.

A small addition to this noise density comes from the op-amp intrinsic
noise.  The total, Spice model\footnote{EECS Department of the
University of California at Berkeley,
http://bwrc.eecs.berkeley.edu/Classes/IcBook/SPICE/} predicted, RMS
width in the current noise density for beam-off and LED off
measurements is $\simeq~19~{\rm fA}/\sqrt{\rm Hz}$~\cite{pp:swb}.

In the DAQ system, the signal from the preamplifier is processed by
the sum and difference amplifiers, where it is fed through a six-pole
Bessel filter with an average $38~{\rm \mu s}$ time constant $(f_{3db}
= 4.16~{\rm kHz})$ (see Section~\ref{scn:DAQ}, Fig.~\ref{fig:SDAS}). The
Bessel filter removes higher frequency noise components present in the
preamplifier output~\cite{pp:swb}. This filtering has to take place
before the digitization. The 16-bit ADCs do not have sufficient
resolution to retain the full noise information at those frequencies
and would cause aliasing of these components into lower
frequencies. Further averaging in the data stream, after digitization,
can not remove this noise without destroying the information content
of the signal.

\begin{figure}[h]
  \hspace{2.5cm}
  \includegraphics[bb=100 0 566 766,clip,angle=270,scale=0.35]{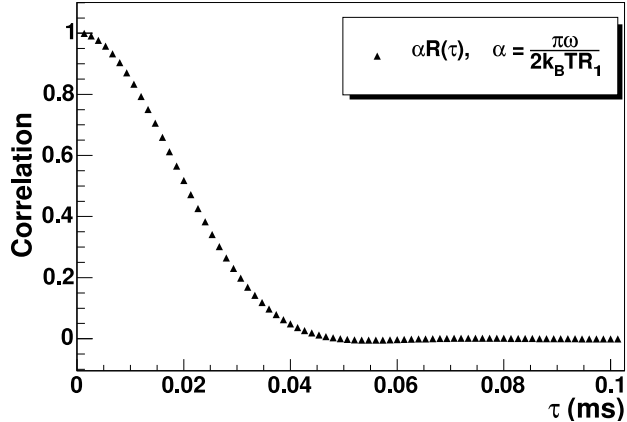}
  \caption{Auto-correlation function for a six pole Bessel filter. The graph shows the correlation 
           between samples taken with a given time difference $(\tau)$ between them. 
	   Here, $\tau = kt_o$ $(t_o = \frac{1}{f_s})$ is the integrated time between the kth sample
	   and the onset of the sampling interval $(P)$, over which samples are taken. At a sampling 
	   rate of either $50$~kHz ($t_o = 0.02$~ms) or $62.5$~kHz ($t_o = 0.016$~ms) the samples are 
	   $\sim 55$\% correlated. The correlation is zero for $\tau > 0.1~{\rm ms}$.} \label{fig:BFACF}
\end{figure}
To find the expected RMS width at the output of the Bessel filter,
one can calculate the corresponding auto-correlation function for the
system~\cite{bb:DVR,pp:bow}.  The effect of the Bessel filter can be
calculated using the corresponding six-pole amplitude response
function~\cite{bb:BLZV,bb:WIND}
\begin{equation}
 \left|\mathcal{H}(i\omega)\right| = \frac{K_o}{\left|B_6(i\omega)\right|} ,
 \label{eqn:PBAC}
\end{equation}

where $B_6(\omega)$ is a 6th order Bessel polynomial and $K_o$ is
chosen such that the dc gain of the filter is unity. The correlation
between time samples can then be calculated {\it via}
\begin{equation}
\mathcal{R}_{I_J}(\tau) =
\int^{\infty}_{0}\left|\mathcal{H}(i\omega)\right|^2~\mathcal{S}_{V_J}(f)~e^{2\pi
if\tau}~df = K_o\int^{\infty}_{0}\frac{4{\rm k_B}TR_1~e^{2\pi
if\tau}~df}{\left|B_6(i2\pi f)\right|^2},\label{eqn:TSCOR}
\end{equation}
whereas the variance seen in an individual sample taken from the
Bessel filter is given by
$$\sigma^2_{k} = \mathcal{R}_{I_J}(\tau=0) =
\int^{\infty}_{0}\frac{4{\rm k_B}TR_1~df}{\left|B_6(i2\pi
f)\right|^2}.$$ The graph shown in Fig.~\ref{fig:BFACF} is the result
of a numerical integration of Eq.~\ref{eqn:TSCOR}, for a range of
intervals $(\tau)$ between samples.

Due to the Bessel filter, the values sampled by the ADCs are highly
correlated at the sampling frequencies of $50$~kHz ($t_o = 0.02$~ms)
or $62.5$~kHz ($t_o = 0.016$~ms) used in the DAQ, indicating that any
high frequency components in the preamplifier output are filtered
out. As a result of this oversampling, the variance for an individual
time bin will be dominated by the correlation between samples.

Given the sampling scheme described in Section~\ref{scn:DAQ}, the
total variance for a time bin in the sum and difference signal is
given by~\cite{bb:DVR}
\begin{equation}
 \sigma^2(V) = \frac{{\sigma_k}^2}{N} + \frac{2~t_o}{P}\sum^{N}_{k=1}\left(1-\frac{k~t_o}{P}\right)\mathcal{R}(k~t_o) .
 \label{eqn:TSCF}
\end{equation}
Here the correlation $\mathcal{R}(\tau = k~t_o)$ is given by
Eq.(\ref{eqn:TSCOR}), $\tau$ is the time difference between the
initial sample and any subsequent sample taken during the sampling
interval $(P)$ and $N$ is the total number of samples taken during the
sampling interval.

In the calculation of the relevant statistic (mean, standard
deviation, etc ...), if we consider an individual time bin measurement
as the fundamental random variable then $P=0.4$~ms, $N = 20$ or $25$,
and $\mathcal{R}(\tau)$ will be the correlation between all ADC
samples within that time bin. On the other hand, if we integrate over
an entire macro pulse and consider that value to be the fundamental
quantity then $P=40$~ms and $\mathcal{R}(\tau)$ will be the
correlation between all ADC samples within that macro pulse and $N =
2000$ or $2500$. According to Fig.~\ref{fig:BFACF} any average value
obtained for one time bin is essentially statistically independent
from the value obtained for any other time bin. Taking either a time
bin or an entire macro pulse as the fundamental quantity is therefore
equivalent, provided that the measured signals have a distribution
that is independent of time over the chosen interval. For beam-on
measurements the signals observed in the detector array have a time
dependence, because the neutron flux from the spallation source varies
with neutron energy. The polarization of the neutrons as well as the
capture distribution in the target are also energy (time) dependent.
So in the measurement of an asymmetry from radiative neutron capture
at a pulsed neutron source, it is necessary to select the time bin
interval as the fundamental statistical quantity.  The $0.4$~ms time
bin width provides an energy resolution small enough to suppress the
corresponding errors below the statistical limit~\cite{pp:greg}.

Using the known values of the preamplifier resistors and the expected
noise density from the first amplifier stage and taking a time bin as
the sampling interval, Eq.(\ref{eqn:TSCF}) gives an expected RMS width
at the preamplifier output of $\simeq~0.1~{\rm mV}$. The bandwidth is
set by the Bessel filter time constants.  For samples taken with $\tau
> 0.05$~ms, the second term in equation~\ref{eqn:TSCF} evaluates to
less than $1~{\rm \mu V}$ and increases by less than 1\% when 25
samples are taken as is done for the sum channels. Gain factors
multiplying the RMS width after the preamplifier are known and the
width observed in the data can be related back to the noise at the
preamplifier output. Other contributions to the noise introduced by
the sum and difference amplifiers are negligible compared to the noise
of the preamplifier.

\subsubsection{Noise Measurement Results}~\label{scn:NMRE}

\begin{figure}[h]
  \hspace{2cm}
  \includegraphics[scale=0.45]{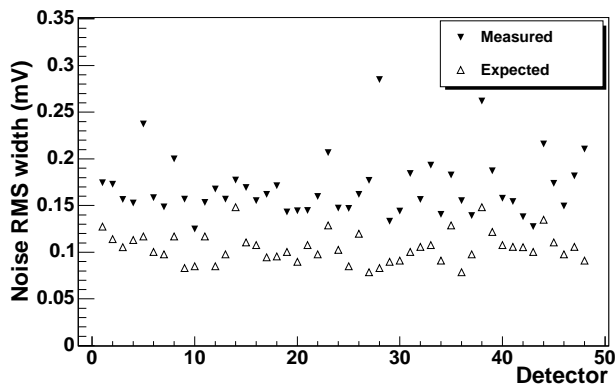}
  \caption{Calculated and measured noise levels for all 48 detectors.
           A 3 sigma cut was placed on the samples in the calculation of
           the RMS noise, to filter most of the cosmic background.
           The measured noise levels include contributions  from any
           activation within the CsI crystals as well as any radiation 
           and electro-magnetic backgrounds found in the general area where 
           the measurement was performed.} \label{fig:JNME}
\end{figure}

The RMS width of the noise is expected to be different for each
detector due to the different feedback resistance values in each
preamplifier (see section~\ref{scn:CRGT}).  The noise was measured
with the data acquisition and the data were averaged over an 8 minute
period. The noise seen in the detectors is shown in
Fig.~\ref{fig:JNME}.

\begin{figure}[h]
  \hspace{1.5cm}
  \includegraphics[scale=0.45]{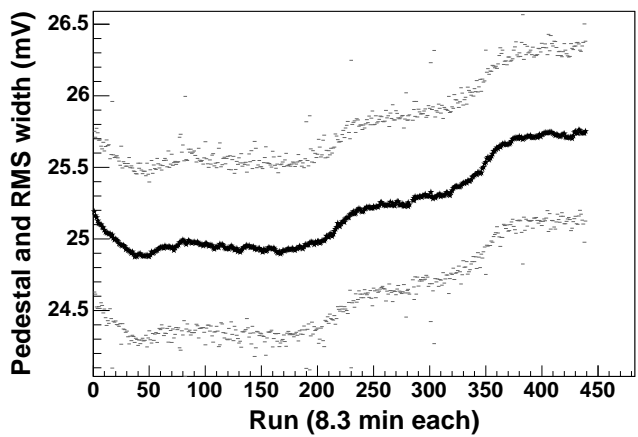}
  \caption{Long-term pedestal and noise vs. run (time) for a typical
           detector. Each run is 8.3 minutes long. The center band is the mean
           run pedestal. The data points above and below the pedestal mean indicate 
	   the noise RMS width.} 
  \label{fig:PEDSTAB}
\end{figure}

Detector pedestals and noise were monitored over a period of 60
hours. The pedestals were seen to drift by about $1$~mV on average and
the noise RMS width stayed the same (Fig.~\ref{fig:PEDSTAB}).

The measured noise levels shown in Fig.~\ref{fig:JNME} also include
contributions from dark currents and cosmic ray background (See
section~\ref{scn:CRB}), even after a 3 sigma cut to filter large
cosmic ray signals. These contributions to the noise are not accounted
for in the noise expected from the calculation above.

The measured RMS width in Fig.~\ref{fig:PEDSTAB}, on the other hand,
includes all contributions to the noise and is observed to be a factor
of 5 higher, on average, than the estimated noise level. These noise
levels determine the time required to perform beam-off measurements of
false asymmetries. During beam-on measurements, the expected shot
noise RMS width is approximately $28$~mV, a factor of $\sim 100$
larger than the largest noise components for beam-off measurements.

\subsection{Cosmic Ray Background}\label{scn:CRB}

\begin{figure}[h]
  \hspace{2.5cm}
  \includegraphics[scale=0.4]{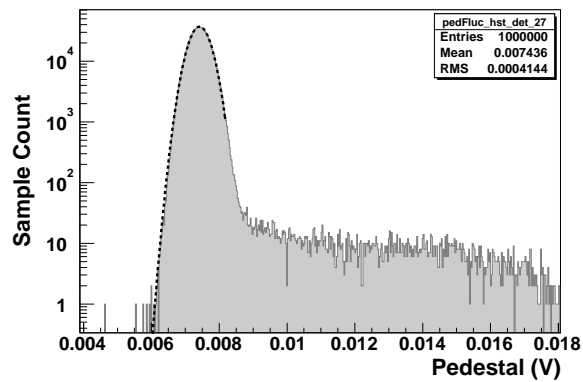} 
 \caption{Electronic pedestal histogram for a typical detector.
           The fit (dashed line) shows an RMS width of $\sim~2.3\times 10^{-4}$~V.
           The data were taken over an 8.3 minute long period.} 
  \label{fig:CRBSIG}
\end{figure}

The detector signals seen during a pedestal run exhibit frequent
sample outliers many standard deviations above the mean (see
Fig.~\ref{fig:DOUT} and Fig.~\ref{fig:CRBSIG}).  These are due to
cosmic rays incident on the detector array.

Considering the surface area of the detector array and its location,
the number of cosmic muons that are expected to enter the array, per
run, is on the order of a few times $10^5$.  This corresponds to
detection of cosmics in less than $0.5$\% of the time bin samples
taken in a given sampling period and a rate of about $7$~Hz in a
single detector.  Several measurements were performed to establish
that the observed outliers do, in fact, correspond to cosmic
radiation. These measurements used filtering techniques similar to
those used on the L3 detector at CERN~\cite{pp:bkk} and the SND
detector at the Budker Institute for Nuclear Physics ~\cite{pp:ach}.

\begin{figure}[h]
  \hspace{2.5cm}
  \includegraphics[bb= 0 0 766 400,clip,scale=0.4]{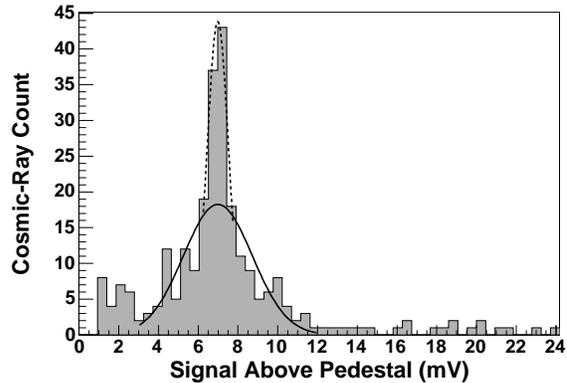}
  \caption{Histogram for muons traversing an entire detector. The detector pedestal 
           has been subtracted. The narrow peak (dashed line) is due to those particles 
           entering the detector normal to the crystal surface. The broader base (solid black
           line) emerges due to particles that enter the crystal at a slight angle
           with respect to its surface normal. These particles have a slightly longer or 
           shorter path length, therefore depositing different amounts of energy.
           The mean of the narrow peak is $\simeq~6.94~\pm~0.09$~mV.} 
  \label{fig:CRBPEAK}
\end{figure}

One of the measurements incorporated the use of cosmic ray event
coincidences between a pair of detectors, one below the other. Two
additional scintillator paddles were installed above and below the
pair to trigger only those muons traversing the entire detector. For
such events, the expected energy deposition is about $100$~MeV.
Figure~\ref{fig:CRBPEAK} shows the result, as obtained for the upper
detector of a pair. According to the measured CsI photoelectron
yield and the known detector gains, an instantaneous energy deposition
of $100$~MeV should produce a signal, in one time bin, of
approximately 7 mV above the pedestal mean.

\subsection{False Asymmetries}\label{scn:FAM}

In this section, we describe measurements performed to study
systematic effects that may introduce false asymmetries as well as the
time required to measure these effects (see Section~\ref{scn:SYSE}).

The first measurement describes the tests for the sensitivity of the
VPDs to AC and DC magnetic fields.  The second measurement was done to
establish the time required to average electronic noise down to the
desired accuracy in the asymmetry. Here we essentially measured the
asymmetry due to electronic noise, which is expected to be zero. This
was done without operating any equipment other than the detector array
itself and the DAQ. For the third measurement the RF spin flipper was
operated and data were taken without any signal going into the
detector array. This was done to search for an additive effect, in
which an asymmetry may be induced as a result of an addition to the
signal in the VPDs, due to spin flipper correlated electronic
pickup. The fourth measurement looked for a multiplicative effect, a
spin flipper correlated gain change in the VPDs, due to any spin
flipper magnetic field leakage. This effect can only be seen by having
a signal (light) going into the VPD, which was accomplished by using
the LEDs in each detector so that the VPD current was approximately
equal to that produced by the scintillation light expected during
beam-on measurements.

\subsubsection{VPD Magnetic Field Sensitivity}\label{scn:VPDBACDC}

VPD gain can change due to magnetic fields interacting with the
photoelectrons.  This nonlinear effect increases with increasing
current between cathode and anode.  As already mentioned in
section~\ref{scn:SYSE}, if this effect is large, small fluctuations in
the field could cause pulse-to-pulse variations in the detector signal
and therefore produce false asymmetries. The sensitivity of the VPDs
to magnetic fields was measured using both dc and ac fields.

An unshielded VPD connected to a preamplifier and a green LED were
placed into a light-tight box which was located in a magnetic field up
to $10$~G . The output of the preamplifier was monitored with a
lock-in amplifier. The VPD was tested in a $10$~G dc field used in the
experiment to control the neutron polarization and suppress
Stern-Gerlach steering. The LED was pulsed at $90$~Hz and produced a
$100$~mV peak-to-peak signal with various offsets up to $1.0$~V at the
preamplifier output.  The tests were performed with the VPD in
parallel and perpendicular to the magnetic field direction as well as
with the VPD rotated around its axis of symmetry. In each
configuration, the change in gain for a dc field is only about
$1\times 10^{-4}{\rm{/G}}$.  No gain dependence on the voltage offset
was observed.

For the ac measurement, the LED was held at a constant voltage
provided by a battery. The magnetic field was varied according to
$B\sin{\omega t}$ with $B \simeq 15$~G. The lock-in amplifier was used
to measure first and second order changes in the gain and was
synchronized to the field frequency.  To first and second order the
gain changes were $2\times 10^{-5}{\rm{/G}}$ and $1\times
10^{-5}{\rm{/G^2}}$ respectively~\cite{pp:greg}. The Aluminum housing
normally placed around the VPD further reduces the coupling and for a
pulse-to-pulse fluctuation of a few mG in the holding field the gain
change in the VPD is negligible.

\subsubsection{Asymmetry Definition}\label{scn:ASYDEF}

Given the vertical polarization direction of the neutrons in the
NPDGamma experiment, the parity violating asymmetry is essentially
seen in a difference of the number of $\gamma$-rays going up and
down. For an asymmetry A, the $\gamma$-ray cross section is
proportional to $1 + A~\cos{\theta}$, where $\theta$ is the angle
between the neutron polarization and the momentum of the emitted
photon.

In calculating a false asymmetry, one asymmetry was calculated for
each time bin and over any valid sequence of eight consecutive macro
pulses (see Section~\ref{scn:DAQ}) with the correct neutron spin state
pattern. A valid 8-step sequence of spin states is defined as
$\uparrow\downarrow\downarrow\uparrow\downarrow\uparrow\uparrow\downarrow$.
This pattern suppresses first and second order gain drifts within the
sequence.
\begin{figure}[h]
  \hspace{3.5cm}
  \includegraphics[scale=0.45]{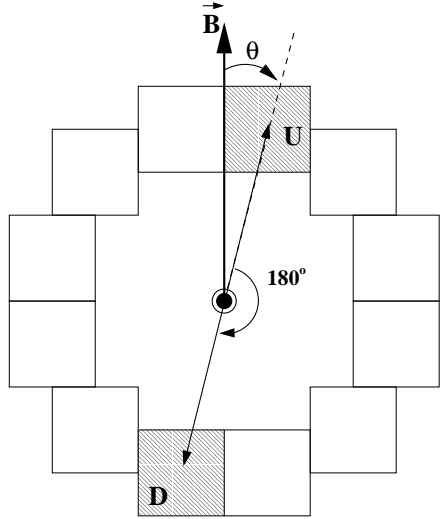}
  \caption{A ring of detectors and one up-down pair, as seen with beam direction into the page. $\vec{B}$ is the magnetic
           holding field defining the direction of the neutron polarization.} \label{fig:RING}
 \end{figure}

A pair of detectors is defined as shown in Fig.~\ref{fig:RING}. If we
let ($U_{\uparrow} , U_{\downarrow}$ or $D_{\uparrow} ,
D_{\downarrow}$) denote the sum of all four signals with the
corresponding spin states in a spin sequence, then the asymmetry for a
time bin is given by

\begin{equation}
  A_{noise} = \frac{1}{d}(U_{\uparrow}-D_{\uparrow}-U_{\downarrow}+D_{\downarrow}).
  \label{eqn:ASY}
\end{equation}

For the LED-on tests and beam-on data the denominator $(d)$ is given
by the sum over all detector signals entering into the numerator of
Eq.(\ref{eqn:ASY}).  For the beam-off, LED-off data the size of the
denominator is set by the expected rate of $\gamma$-rays when the beam
is on, as well as gain and sampling factors in the DAQ.

\subsubsection{False Asymmetry Results} \label{scn:FAR}
For the spin flipper on and off (no LED) asymmetry measurements, the
time required to achieve a certain accuracy in the asymmetries is
limited by the RMS width in the signal.  As described above, the RMS
width is set by the Johnson noise Eq.(\ref{eqn:TSCF}) and additional
noise from the detector preamplifier~\cite{pp:swb} as well as cosmic
ray and other background. The data were analyzed both with cuts to
remove most of the cosmic-ray background and without cuts to study the
influence of cosmic-rays on false asymmetries.  The cuts were placed
on the individual time bin samples and excluded samples larger than
six times the noise width (6 sigma) seen in the pedestals.

\begin{figure}[h]
 \hspace{2.5cm}
  \includegraphics[scale=0.45]{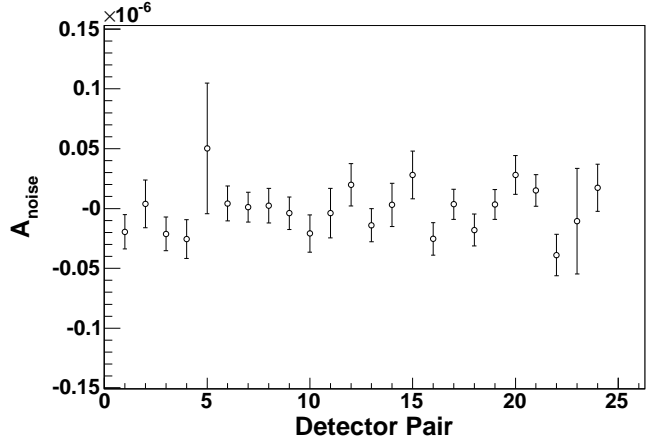}
  \caption{Measured additive false asymmetries with the spin flipper on and LEDs off.} \label{fig:ASYSP}
 \end{figure}

Many data runs contribute to the asymmetry measurements and for each
detector pair in the array a combined mean and standard deviation were
calculated from all runs.  A total error-weighted average asymmetry is
then calculated for the entire array. Without LEDs, the asymmetry can
be measured down to the $5\times10^{-9}$ level in 3 hours. In 2 hours,
the additive false asymmetry with the spin flipper was measured to
$$
A_{noise} = \left(-4 \pm  3\right)\times10^{-9}.
$$ The individual detector pair asymmetries from the spin flipper runs
are shown in Fig.~\ref{fig:ASYSP}.  The noise observed in the
detectors, with the spin flipper running, did not significantly change
from the levels shown in Fig.~\ref{fig:JNME}. If no cuts are applied
to remove the cosmic background, the time required to measure the
noise asymmetry to the above accuracy increases by about a factor of
five and the asymmetry is still consistent with
zero. Table~\ref{tbl:ANOISE} shows the spin flipper on, LED off
asymmetries for each ring in the detector array.

\begin{table}[h]
  \begin{center} \label{tbl:ANOISE}
    \begin{tabular}{|l|c|c|l|c|c|} \hline
                      & $A_{noise}$                           \\ \hline \hline
      Ring 1          & $\left(-12 \pm  7\right)\times10^{-9}$\\ \hline
      Ring 2          & $\left(-1  \pm  6\right)\times10^{-9}$\\ \hline
      Ring 3          & $\left(-7  \pm  6\right)\times10^{-9}$\\ \hline
      Ring 4          & $\left( 6  \pm  7\right)\times10^{-9}$\\ \hline
    \end{tabular} 
    \caption{Spin Flipper on, LED off Additive False Asymmetry by Detector Ring.}
  \end{center}
\end{table}
  
\begin{figure}[h]
 \hspace{2.5cm}
  \includegraphics[scale=0.45]{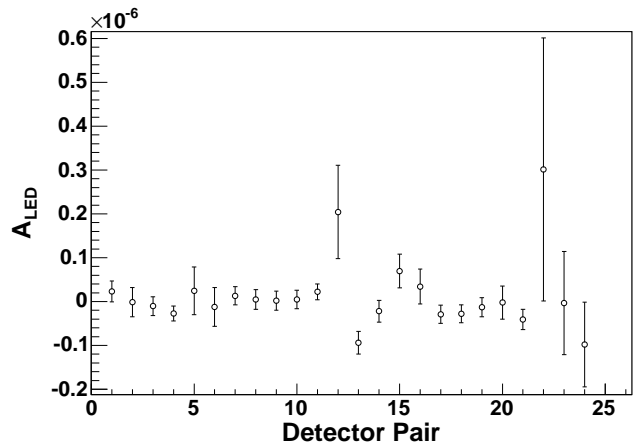}
  \caption{Measured multiplicative false asymmetries with the spin flipper and LEDs on.
           The large errors on some of the pair asymmetries are due to noisy LEDs.} \label{fig:ASYSPLED}
 \end{figure}

With the LEDs turned on, the RMS width of the detector signals is
dominated by shot noise at the photocathode. If the shot noise is
characterized by a single electron, then the expected average noise
density is $\simeq~95~{\rm fA}/\sqrt{\rm Hz}$ at an average current of
$28$~nA out of the VPDs. The multiplicative false asymmetry for the 24
detector pairs with LED signal is shown in Fig.~\ref{fig:ASYSPLED}.

In 11 hours, the noise asymmetry with the LED signal, for the combined
array, was measured to
$$
A_{LED} = \left(-1 \pm  4\right)\times10^{-9}.
$$

\begin{table}[h]
  \begin{center} \label{tbl:MNOISE}
    \begin{tabular}{|l|c|c|l|c|c|} \hline
                      & $A_{noise}$                           \\ \hline \hline
      Ring 1          & $\left(-9  \pm   9\right)\times10^{-9}$  \\ \hline
      Ring 2          & $\left( 11 \pm   6\right)\times10^{-9}$ \\ \hline
      Ring 3          & $\left(-14 \pm  10\right)\times10^{-9}$\\ \hline
      Ring 4          & $\left(-16 \pm  13\right)\times10^{-9}$\\ \hline
    \end{tabular} 
    \caption{Spin Flipper on, LED on Multiplicative False Asymmetry by Detector Ring.}
  \end{center}
\end{table}

The noise in the detector preamplifier, as expected from calculation,
predicts a run time estimate of $1$~hour to measure the beam-off,
LED-off noise asymmetry to $5\times10^{-9}$. From the estimate of the
shot noise for LED-on measurements, the run time should be $5$~hours
for an average photo-current of $28$~nA . The performed measurements
show that the required run-times are actually 2 to 3 times larger than
predicted. However, the average beam-off noise in the preamplifier, as
seen in Fig.~\ref{fig:JNME}, is $\simeq 32~{\rm fA}/\sqrt{\rm Hz}$,
about 70\% higher than expected and the runtime required with this
noise level corresponds to $\simeq 3$~hours.  The same was seen to be
true for the LED-on (beam-on) measurement, where the noise observed
was about 50\% larger than expected from the above shot noise
estimate.

\section{Summary} \label{scn:CONC}
The NPDGamma CsI(Tl) detector array has been designed and built to
operate in current mode and at low noise without introducing
instrumental systematic effects at the $10^{-9}$ level.  The
prerequisite for a successful current mode measurement is the
suppression of noise levels much below the statistical limit. The
noise in the preamplifier due to thermal fluctuations in the circuit
components was measured to be smaller, by a factor of 70, than the
expected shot noise during measurements with a neutron beam. Since, in
current mode detection, counting statistics appear as shot noise at
the VPD photo-cathode, the accuracy of the asymmetry measurement will
therefore be determined by counting statistics.

Any systematic false asymmetries arising due to the operation of the
array and spin flipper must be suppressed to have an effect below
$5\times 10^{-9}$.  The beam-off, LED-off additive false asymmetry due
to spin flipper correlated electronic pickup as well as the LED-on
multiplicative false asymmetry due to spin flipper correlated gain
changes in the VPD were measured to this level of accuracy within a
few hours and were consistent with zero. These results show that the
detector array is operating as designed and meets all criteria needed
to perform a successful measurement of the weak parity-violating
$\gamma$ asymmetry with an accuracy of $5\times 10^{-9}$ in the
neutron capture reaction \npdg.

\section{Acknowledgments}
The authors would like to thank Mr.\ G.\ Peralta (LANL) for his
technical support during this experiment, Mr.\ W.\ Fox (IUCF) and
Mr.\ T.\ Ries (TRIUMF) for the mechanical design of the array and the
construction of the stand and Mr.\ M.\ Kusner of Saint-Gobain in Newbury, 
Ohio for interactions during the manufacture and characterization of
the CsI(Tl)crystals. We would also like to thank TRIUMF for
providing the personnel and infrastructure for the stand construction.
This work was supported in part by the U.S.\ Department of Energy
(Office of Energy Research, under Contract W-7405-ENG-36), the
National Science Foundation (Grant No. PHY-0100348) and 
the NSF Major Research Instrumentation program (NSF-0116146), the Natural
Sciences and Engineering Research Council of Canada and the Japanese
Grant-in-Aid for Scientific Research A12304014.

\pagebreak



\begin{thebibliography}{00}
\bibitem{pp:FPPNB}
C. R. Gould, G. L. Greene, F. Plasil, and W. M. Snow, editors,
Fundamental Physics with Pulsed Neutron Beams, World Scientific, ISBN
981-02-4667-6 (2001).

\bibitem{pp:NIST1}
W. M. Snow, In the proceedings of the International Conference on Precision Measurements with
Slow Neutrons, NIST Journal of Research (2004).

\bibitem{pp:NIST2}
S. A. Page, {\it et al.}, In the proceedings of the International Conference on Precision 
Measurements with Slow Neutrons, NIST Journal of Research (2004).

\bibitem{pp:NIST3}
A. Komives, In the proceedings of the International Conference on Precision Measurements with
Slow Neutrons, NIST Journal of Research (2004).

\bibitem{pp:prop}
J. D. Bowman, {\it et al.}, Measurement of the parity-violating gamma asymmetry ${\rm A_{\gamma}}$
in the capture of polarized cold neutrons by para-hydrogen, \npdg~, Tech. Rep. LA-UR-99-5432,
Los Alamos National Laboratory (1999).

\bibitem{pp:snow}
W. M. Snow, {\it et al.}, Nucl. Instr. Meth. A 440, (2000) 729.

\bibitem{pp:snow2}
W. M. Snow, {\it et al.}, Nucl. Instr. Meth. A 515, (2003) 563.

\bibitem{pp:DDH}
B. Desplanques, J. F. Donoghue, B. R. Holstein, Annals of Physics 124, (1980) 449.

\bibitem{pr:MCNP}
J. F. Briesmeister, ed., MCNP, A General Monte Carlo N-Particle 
Transport Code, Version 4C, LA-13709-M, Los Alamos National Laboratory (2000).

\bibitem{pr:EGS4}
W. R. Nelson, H. Hirayama, D. W. O. Rogers, The EGS4 Code System, SLAC 265, SLAC, (1985).

\bibitem{rep:Macfarlane}
R. E. MacFarlane, Technical Report No. LA-12146-C, Los Alamos National Laboratory
(unpublished).

\bibitem{pp:pil}
P.-N. Seo, {\it et al.}, Nucl. Instr. Meth. A 517 (2004) 285. 

\bibitem{pp:Grassman}
H. Grassman, {\it et al.}, Nucl. Instr. Meth. 228 (1985) 323. 

\bibitem{pp:Schotanus}
P. Schotanus, {\it et al.}, IEEE Trans. Nucl. Sci. 37 (1990) 177. 

\bibitem{pp:FRLEZ}
E. Frlez, B. K. Wright, D. Pocanic, Optics: General-Purpose Scintillator Light
Response Simulation Code, Computer Physics Communications 0 (2000), 1-26.

\bibitem{bb:Knoll}
G. F. Knoll, Radiation Detection and Measurement, John Wiley and Sons,
New York, 1989.

\bibitem{pp:REN}
D. Renker, Proceedings of the ECFA Study Week on Instrumentation
for High-Luminosity Hadron Collider, (CERN, Geneva, 1989), Vol. CERN 89-10.

\bibitem{pp:ZHU}
R. Zhu, Nucl. Instr. Meth. A 413 (1998) 297. 

\bibitem{pp:WEI}
Z. Wei and R. Zhu, Nucl. Instr. Meth. A 326 (1993) 508. 

\bibitem{pp:CHD}
M. A. H. Chowdhury and D. C. Imrie, Nucl. Instr. Meth. A 432 (1999) 138. 

\bibitem{pp:KOB}
M. Kobayashi and S. Sakuragi, Nucl. Instr. Meth. A 254 (1987) 275. 


\bibitem{pp:greg}
G. S. Mitchell, {\it et al.}, Nucl. Instr. Meth. A 521 (2004) 468.

\bibitem{pp:swb}
W. S. Willburn, J. D. Bowman, M. T. Gericke, S. I. Penttil{\"a}, 
submitted to Nucl. Instr. Meth. A (2004).

\bibitem{pp:pmsn}
M. T. Gericke, {\it et al.}, In the proceedings of the International Conference on Precision 
Measurements with Slow Neutrons, NIST Journal of Research (2004).

\bibitem{bb:DVR}
W. B. Davenport, W. L. Root, An Introduction to the Theory of Random Signals and Noise,
John Wiley \& Sons, New York, 1987.

\bibitem{bb:HWHL}
P. Horowitz, W. Hill, The Art of Electronics,
Cambridge University Press, United Kingdom, 2001.

\bibitem{pp:CGP}
A. Csoto, B. F. Gibson and G. L. Payne, Phys. Rev. C 56 (1997) 631.

\bibitem{pc:Ino}
T. Ino, {\it et al.}, Private communication.

\bibitem{pp:bow}
J. D. Bowman and J. C. VanderLeeden, Nucl. Instr. Meth. 85 (1970) 19.

\bibitem{bb:BLZV}
H. J. Blinchikoff, A. I. Zverev, Filtering in the Time and Frequency Domains,
John Wiley \& Sons, New York, 1976.

\bibitem{bb:WIND}
S. Winder, Filter Design, NEWNES Butterworth-Heinemann, Oxford, 1997.

 

\bibitem{pp:bkk}
J. A. Bakken, {\it et al.}, Nucl. Instr. Meth. A 275 (1989) 81.

\bibitem{pp:ach}
M. N. Achasov, {\it et al.}, Nucl. Instr. Meth. A 401 (1997) 179.

\end{thebibliography}

\end{document}